\documentclass[prd,twocolumn,showpacs,preprintnumbers,amsmath,amssymb]{revtex4}
\usepackage{cancel}
\usepackage{booktabs}
\usepackage{multirow}
\usepackage{slashed}
\usepackage{graphicx}
\usepackage{array}
\usepackage{color}
\usepackage[colorlinks,
            citecolor=blue,
            anchorcolor=blue,
            menucolor=blue,
            linkcolor=red,
            filecolor=blue,
            runcolor=red,
            urlcolor=blue,
            frenchlinks=red]{hyperref}
\usepackage{epstopdf}
\begin{document}

\title{Constructing new pseudoscalar meson nonets with the observed $X(2100)$, $X(2500)$, and $\eta(2225)$}

\author{Li-Ming Wang$^{1,2}$}\email{lmwang15@lzu.edu.cn}
\author{Si-Qiang Luo$^{1,2}$}\email{luosq15@lzu.edu.cn}
\author{Zhi-Feng Sun$^{1,2}$}\email{sunzhif09@lzu.edu.cn}
\author{Xiang Liu$^{1,2}$\footnote{Corresponding author}}
\email{xiangliu@lzu.edu.cn}
 \affiliation{
$^1$School of Physical Science and Technology, Lanzhou University, Lanzhou 730000, China\\
$^2$Research Center for Hadron and CSR Physics, Lanzhou University
and Institute of Modern Physics of CAS, Lanzhou 730000, China }

\begin{abstract}
Stimulated by the BESIII observation of $X(2100)$, $X(2500)$, and
$\eta(2225)$, we try to pin down new pseudoscalar meson nonets
including these states. { The analysis of mass spectra
and the study of strong decays} indicate that $X(2120)$ and
$\eta(2225)$ associated with $\pi(2070)$ and the predicted kaon
$K(2150)$ may form a new pseudoscalar meson nonet. In addition, more
experimental data for $X(2100)$ { are} necessary to
determine its structure { of nonets}. Then, $X(2500)$,
$X(2370)$, $\pi(2360)$, and the predicted kaon $K(2414)$ can be
grouped into another new nonet. These assignments to {
the} discussed pseudoscalar states can be further tested in
experiment.

\end{abstract}
\date{\today}
\maketitle

%%%%%%%%%%%%%%%%%%%%%%%%%%%%%%%%%%%%%%%%%%%%%%%%%%%%%%
\section{Introduction}\label{sec1}
%%%%%%%%%%%%%%%%%%%%%%%%%%%%%%%%%%%%%%%%%%%%%%%%%%%%%%

In the pseudoscalar meson family, the first nonet is constructed by
$\pi$, $\eta(548)$, $\eta^\prime(958)$, and $K(494)$, and then the
second nonet appears with { the components} of $\pi(1300)$,
$\eta(1295)$, $\eta(1475)$, and $K(1460)$. { As
indicated} in Ref. \cite{Yu:2011ta}, the $X(1835)$ observed in the
$\eta^\prime \pi^+\pi^-$ invariant mass spectrum of $J/\psi\to
\gamma\eta^\prime \pi^+\pi^-$ associated with the $\eta(1760)$,
$\pi(1800)$, and $K(1830)$ forms the third pseudoscalar meson nonet.
By this way, one can categorize well the observed pseudoscalar states
into pseudoscalar meson families. Obviously, this is not the end of the
{ whole} story.

In 2016, the BESIII Collaboration \cite{Ablikim:2016hlu} performed a
partial wave analysis of the $J/\psi\to \gamma\phi\phi$ decay, by
which two isoscalar { and pseudoscalar}  states $X(2100)$
(which { was} named as $\eta(2100)$ in Ref.
\cite{Ablikim:2016hlu}) and $X(2500)$ were observed with $22\sigma$
significance and $8.8\sigma$ significance, respectively. In
addition, $\eta(2225)$, which was first reported in Ref.
\cite{Ablikim:2008ac}, was confirmed with $28\sigma$ significance.
Their corresponding resonance parameters were measured as
\cite{Ablikim:2016hlu}
\begin{eqnarray}
m_{X(2100)}&=&2050^{+30+75}_{-24-26}\,{\text{MeV}},\\
\Gamma_{X(2100)}&=&250^{+36+181}_{-30-164}\,{\text{MeV}},\\
m_{X(2500)}&=&2470^{+15+101}_{-19-23}\, {\text{MeV}},\\
\Gamma_{X(2500)}&=&230^{+64+56}_{-35-33}\, {\text{MeV}},\\
m_{\eta(2225)}&=&2216^{+4+21}_{-5-11} \,{\text{MeV}},\\
\Gamma_{\eta(2225)}&=&185^{+12+43}_{-14-17}\, {\text{MeV}}.
\end{eqnarray}

These newly observed $X(2100)$, $X(2500)$, and $\eta(2225)$ provide us a good chance to construct new pseudoscalar meson nonets with higher radial excitations. Mainly considering this point, in this work, we study whether the newly observed $X(2100)$, $X(2500)$ and $\eta(2225)$ can be categorized into pseudoscalar meson nonets. First, we perform an analysis of the Regge trajectories, which provides an important hint of how to group these pseudoscalar states into new pseudoscalar meson families. Second, we study their two-body strong decays by the flux-tube model, which can be applied to test the possible assignments. In the following sections, we will give detailed illustrations.

When constructing pseudoscalar meson nonets with higher radial excitations, the corresponding pseudoscalar kaons are still missing in experiment. Thus, as { an} important theoretical prediction, the masses and decay behaviors of kaons in constructing the nonets will be given, which { may provide valuable information} for a future experimental search for those kaons.

This paper is organized as follows. After the Introduction, we concisely review the research status of the reported pseudoscalar states above 2 GeV in Sec. \ref{sec2}. Then, we present a mass spectrum analysis by the approach of the Regge trajectories in Sec. \ref{sec3}. The two-body decay behaviors of the discussed pseudoscalar states are  given in Sec. \ref{sec4}. The paper ends with the short summary.

\section{concise review of the reported pseudoscalar states above 2 GeV}\label{sec2}

Before the BESIII's analysis, several isoscalar and pseudoscalar states were reported \cite{Yu:2011ta,Li:2008et,Bugg:1999jc,Anisovich:2000ut,Chen:2011kp,Ablikim:2016hlu,Liu:2010tr,Pan:2016bac,Ablikim:2010au,Bisello:1988as,Deng:2012wj,Wang:2010vz}, which include $\eta(2010)$, $\eta(2100)$, $\eta(2190)$, $\eta(2320)$, $X(2120)$, and $X(2370)$. However, these states are not listed in summary meson tables of the Particle Data Group (PDG) \cite{Olive:2016xmw} since they are not confirmed by other experiments. It also means that these states are not established in experiment, either.
The $\eta(2010)$ with mass $2010_{-60}^{+35}$ MeV and width $270 \pm 60$ MeV was found by analyzing $p\overline{p}$ annihilation into $\eta \pi^0 \pi^0$, $\pi^0 \pi^0$, $\eta \eta$, and $\pi^- \pi^+$ \cite{Anisovich:2000ut}.
The $\eta(2100)$ was observed by the DM2 experiment in the radiative decay $J/\psi\to \gamma\rho\rho$ \cite{Bisello:1988as}.
In Ref. \cite{Bugg:1999jc}, the $\eta(2190)$ was introduced by
studying the data of the radiative decays of $J/\psi$ into the $0^-$ final states \cite{Anisovich:2005wf}, which has mass $2190\pm50$ MeV and width $850\pm 100$ MeV.
In Ref.  \cite{Li:2008et}, the authors discussed the possibility of the $\eta(2010)$ and $\eta(2190)$ as $4^1S_0$ isoscalar states. The $\eta(2225)$ was suggested to be a good candidate for the $4^1S_0 \; s\bar{s}$ state \cite{Li:2008we}.
The $\eta(2100)$ and $\eta(2225)$ were treated as the third radial excitations of $\eta$ and $\eta^\prime$, respectively, in Ref. \cite{Li:2008et}. The $\eta(2320)$ was discovered from the combined analysis of $p\bar{p}\rightarrow \eta\eta\eta$ and $p\bar{p} \rightarrow \eta\pi\pi$ \cite{Anisovich:2011ha}.
The $X(2120)$ and $X(2370)$ are two pseudoscalar states observed by BESIII in the { invariant} mass spectrum of the $J/\psi\to \eta^\prime \pi^+\pi^-$ decay \cite{Ablikim:2010au}. The observation of the $X(2120)$ and $X(2370)$ also stimulated the discussions on pseudoscalar meson, glueball, and hadronic molecular state \cite{Liu:2010tr,Yu:2011ta,Chen:2011kp,Wang:2010vz}. In Ref. \cite{Wang:2010vz}, the author has studied the mass spectrum of a baryonium with the Bethe-Salpeter equation \cite{Geng:2016pmf,Windisch:2016iud,Higashijima:1975hf,Falkensteiner:1980xu,Tomozawa:1981da}, and $X(2370)$ can be identified as a $p\bar{N}(1400)$ bound state.

Besides the isoscalar and pseudoscalar states mentioned above, there are two isovector and pseudoscalar states $\pi(2360)$ and $\pi(2070)$ above 2 GeV, which were observed by the Crystal Barrel experiment, where the partial wave analysis of the decay $p\overline{p}\rightarrow \eta \eta \pi^0$ was done \cite{Anisovich:2001pp}. The $\pi(2360)$ has mass $M=2360\pm25$ MeV and width $\Gamma = 300_{-50}^{+100}$ MeV, while the $\pi(2070)$ has mass $M = 2070 \pm 35$ MeV and width $\Gamma = 310_{-50}^{+100}$ MeV. Anisovich {\it et al}. suggested $\pi(2360)$ and $\pi(2070)$ as the third and fourth radial excitations of the $\pi$ meson family, respectively \cite{Anisovich:2005dt}. In Ref \cite{Li:2008et}, the two-body strong decays of $\pi(2070)$ are calculated by the quark pair creation model assuming $\pi(2070)$ as $\pi(4S)$. The $\pi(2360)$ as $\pi(5^1S_0)$ was supported by the analysis of Regge trajectories \cite{Pan:2016bac}.

From this concise review of the observed pseudoscalar states above 2 GeV, we can learn that the experimental and theoretical studies are still in chaos, especially for the isoscalar and pseudoscalar states. In the following, we try to establish new pseudoscalar meson nonets with higher radial excitations by combining the newly observed $X(2100)$, $X(2500)$, and $\eta(2225)$ with the pseudoscalar states already reported.

%%%%%%%%%%%%%%%%%%%%%%%%%%%%%%%%%%%%%%%%%%%%%%%%%%%%%%%%%%%%%
\section{Mass spectrum analysis}\label{sec3}
%%%%%%%%%%%%%%%%%%%%%%%%%%%%%%%%%%%%%%%%%%%%%%%%%%%%%%%%%%%%%
In the light pseudoscalar sector, $\eta$ and $\eta^\prime(958)$
together with $\pi$ and $K$ can be elements of the lowest meson nonet,
while $\eta(1475), \eta(1295), K(1460)$, and $\pi(1300)$ form the
meson nonet with the first radial excitation. In Refs.
\cite{Yu:2011ta,Li:2008mza,Huang:2005bc,Wang:2010vz,Anisovich:2005dt,Liu:2010tr},
$X(1835),\eta(1760),K(1830)$, and $\pi(1800)$ are depicted as the
states with quantum number $3^1S_0$. What we will discuss in this
paper {is} the third and fourth radial excitations
of { pseudoscalar} mesons.

The Regge trajectory analysis \cite{Chew:1962eu,Kovalenko:2015ssa} is a
practical way to study the mass spectrum
\cite{Page:1995rh,Capstick:1986bm,Capstick:1993kb,Ackleh:1996yt} of
mesons. The relation between mass and the radial quantum number $n$ is
\begin{equation}\label{equa7}
M^2=M_0^2+(n-1)\mu^2,
\end{equation}
where $M_0$ and $M$ are the masses of ground state and $(n-1)$th
radial excitation state, respectively. $\mu^2$ denotes the slope of the trajectory with
the value $\mu^2=1.25\pm0.15\,$GeV$^2$ \cite{Anisovich:2000kxa}.

\begin{figure*}[htbp]
\begin{center}
\includegraphics[width=1\textwidth,scale=0.42]{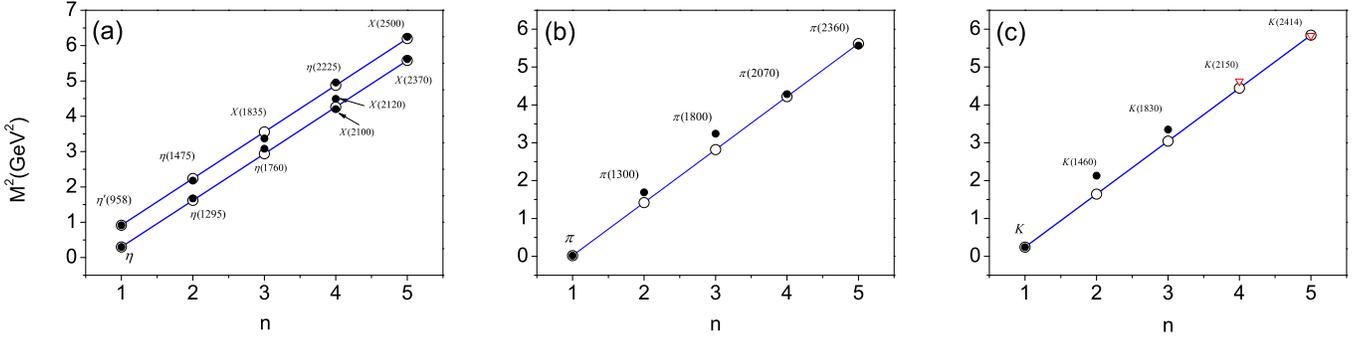}
\caption{The Regge trajectories for the $\eta/\eta^\prime$, and $\pi$ and $K$ mass spectrum with $\mu^2=1.32$, $1.40$, $1.40$ $\rm{GeV}^2$, respectively. Here, $\circ$ denotes Regge trajectories theoretical values. $\triangledown$ denotes theoretical values from Gell-Mann-Okubo mass formula \cite{Li:2008et}. And $\bullet$ denotes experimental values.
}\label{fig1}
\end{center}
\end{figure*}

{When we plot the Regge trajectories in Fig. \ref{fig1}, we notice that
$\pi(2070)$ and $\pi(2360)$ as well as $\pi$, $\pi(1300)$, and $\pi(1300)$ populate a common trajectory.} For $\eta(4S)$, the predicted mass by
the analysis of Regge trajectories is about 2064 MeV, where
$\eta(2010)$, $\eta(2100)$, $\eta(2190)$, $X(2100)$, and $X(2120)$
are its candidates. Similarly, $X(2370)$, $\eta(2225)$, and $X(2500)$
are candidates for $\eta(5S)$, $\eta^\prime(4S)$, and
$\eta^\prime(5S)$, respectively. The former theoretical studies on
the {masses} of $\pi(4S)$, $\eta(4S)$, $\eta^\prime(4S)$,
$\pi(5S)$, $\eta(5S)$, and $\eta^\prime(5S)$
\cite{Krasnikov:1981vw,Gorishnii:1983zi,Hatanaka:2008xj,Anisovich:2005dt,Shu:2016rrr,GolecBiernat:2007xv,Anisovich:2000kxa,
Li:2008et, Pan:2016bac} are consistent with the trajectory analysis
in our work.

For the sake of completeness, the kaons with higher radial excitation
should appear in the corresponding nonets. However, there is no
experimental information about them with quantum numbers
$4^1S_0$ and $5^1S_0$. With the help of diagonalization of the mass
squared matrix and Gell-Mann-Okubo mass formula, the following
relation is obtained \cite{Li:2008et},
\begin{eqnarray}\label{equa8}
&&8X^2(M^2_{K(n^1S_0)}-M^2_{\pi(n^1S_0)})^2\nonumber\\
&&=\big[4M^2_{K(n^1S_0)}-(2-X^2)M^2_{\pi(n^1S_0)}-(2+X^2)\nonumber\\
&&\quad\times M^2_{X(n^1S_0)}\big]
\big[(2-X^2)M^2_{\pi(n^1S_0)}+(2+X^2)\nonumber\\
&&\quad\times M^2_{X(n^1S_0)}-4M^2_{K(n^1S_0)}\big],
\end{eqnarray}
where $X$ describes the SU(3)-breaking ratio of the nonstrange and
strange quark propagators via the constituent quark mass ratio
$m_u/m_s$. The masses of $K(4^1S_0)$ and $K(5^1S_0)$ are predicted to be
2150 and 2414 MeV, respectively, so that we label them as $K(2150)$ and $K(2414)$,
respectively. In addition, these two states are approximately located on
the trajectory for kaons.

The mass information only is not sufficient to classify the
structure of the states mentioned above. So, we study their two-body
strong decay in the next section.

%%%%%%%%%%%%%%%%%%%%%%%%%%%%%%%%%%%%%%%%%%%%%%%%%%%%%%%%%%%%%
\section{Two-body strong decay behaviors}\label{sec4}
%%%%%%%%%%%%%%%%%%%%%%%%%%%%%%%%%%%%%%%%%%%%%%%%%%%%%%%%%%%%%

\subsection{Brief introduction of the flux-tube model}\label{subA}

In this section, we study the strong decay behaviors of the third and
fourth radial excited pseudoscalar meson nonets by the flux-tube model
\cite{Isgur:1984bm,Blundell:1996as,Zhao:2005nu,Deng:2012wj,Li:2008xy,Geiger:1994kr}.
In the following, a brief introduction of this model is given.

\begin{figure}[hbtp]
\centering
\includegraphics[width=0.42\textwidth,scale=0.42]{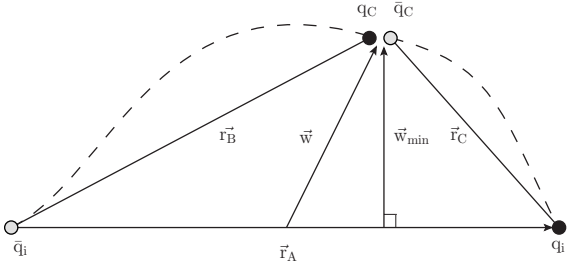}
\caption{The position-space coordinates used in the flux-tube model. }\label{fig2}
\end{figure}

The flux-tube model, first proposed by Isgur and Paton, is a useful
tool for describing the decay properties of hadrons. It is suggested
by the strong coupling limit of the Hamiltonian lattice QCD. In this
model, { a quark and an antiquark compose a meson} and are connected by
a chromoelectric flux tube. Here, the flux tube can be treated as
a vibrating string. Figure \ref{fig2} describes the picture of a meson
decay, which happens when the string breaks at a point, and then the
free ends of the flux tube for an initial meson (i.e., $q_i$ and
$\bar{q}_i$) connect with the quark-antiquark pair { ($q_C$ and
$\bar{q}_C$)} created from the
vacuum.
\\

In this paper, within the frame of the flux-tube model, the expression
of a partial wave amplitude is
\begin{equation}\label{equa9}
\begin{array}{l}
\mathcal{M}^{SL}(P)=\gamma_0
\frac{\sqrt{32\pi(2L+1)E_AE_BE_C}}{2J_A+1}\;\sum\limits_{
\scriptscriptstyle{\substack{ M_{L_A},M_{S_A},M_{L_B}, M_{S_B},\\ M_{L_C},M_{S_C},M_{J_B},M_{J_C},m}}}\\
\times\langle L0S(M_{J_B}+M_{J_C})|J_A(M_{J_B}+M_{J_C}) \rangle\\
\times\langle J_B M_{J_B} J_C M_{J_C} |S (M_{J_B}+M_{J_C}) \rangle\\
\times\langle L_A M_{L_A} S_A M_{S_A} |J_A(M_{J_B}+M_{J_C}) \rangle\\
\times\langle L_B M_{L_B} S_B M_{S_B} |J_B M_{J_B} \rangle \langle L_C M_{L_C} S_C M_{S_C} |J_C M_{J_C} \rangle\\
\times\langle 1m1-m|00\rangle\; \langle \chi_{S_B M_{S_B}}^{14}\chi_{S_C M_{S_C}}^{32}|\chi_{S_A M_{S_A}}^{12}\chi_{1-m}^{34}\rangle\\
\times \bigl[\langle \phi_B^{14}\phi_C^{32}|\phi_A^{12}\phi_0^{34}\rangle\; I^{\textrm{ft}}(P \vec{e}_z,m_1,m_2,m_3)\\
+(-1)^{L_A+L_B+L_C+S_A+S_B+S_C}\langle\; \phi_B^{32}\phi_C^{14}|\phi_A^{12}\phi_0^{34}\rangle\\
\times I^{\textrm{ft}}(P \vec{e}_z,m_2,m_1,m_3) \bigr].
\end{array}
\end{equation}
{ Here, $P$ is the momentum of a meson $B$. $S$
and $L$ denote the total spin and relative orbital angular momentum
between mesons $B$ and $C$, respectively. $E_B$ is the total energy
of a meson $B$. $L_a$ is the orbital angular momentum between a quark
and antiquark in a meson $a$ ($a=A, B, C$). $J_a$ is the total spin of
$a$. $M_{L_a}$ and $M_{J_a}$ are the magnetic quantum numbers
corresponding to $L_a$ and $J_a$. $m_1$ and $m_2$ are quark masses
in a meson A; $m_3$ is the mass of the quark and antiquark created from the vacuum.
$\chi_{s_a,m_{s_a}}^{ij} $ and $\phi_a^{ij}$ are spin and flavor
wave functions of quark $i$ and $j$, and $\phi_0^{ij}$ is the flavor
wave function of the quark and antiquark created from the vacuum.} The
space integral of the last factor in Eq.(\ref{equa9}) is given as follows,
\begin{equation}\label{equa10}
\begin{array}{l}
I^{\textrm{ft}}\left(P \vec{e}_z,m_1,m_2,m_3\right)=-\frac{8}{(2\pi)^{3/2}}\displaystyle{\int d^3\vec{r}\int d^3\vec{w}}\\
\times \psi_{n_B L_B M_{L_B}}^\ast(-\vec{w}-\vec{r})\psi_{n_C L_C M_{L_C}}^\ast(\vec{w}-\vec{r})\\
\times y_1^m\left(\left[(P \vec{e}_z+i\vec{\triangledown}_{\vec{r}_A})\psi_{n_A L_A M_{L_A}}(\vec{r}_A)\right]_{\vec{r}_A=-2\vec{r}}\right)\\
\times
\textrm{exp}(-\frac{1}{2}bw_{\textrm{min}}^2)\;\textrm{exp}\big(i\vec{P}\cdot(m_+\vec{r}+m_-\vec{w})\big),
\end{array}
\end{equation}
with $m_+=\frac{m_1}{m_1+m_3}+\frac{m_2}{m_2+m_3}$,
$m_-=\frac{m_1}{m_1+m_3}-\frac{m_2}{m_2+m_3}$. { The quark pair
creation (QPC) model \cite{Geng:2016pmf} was first proposed by Micu to calculate
Okubo-Zweig-Iizuka (OZI) strong decays. In the QPC model, the heavy flavor meson decay
occurs through a quark-antiquark pair production from the vacuum,
which has the quantum number of the vacuum, i.e., $0^{++}$. In
the QPC model, the constant $\gamma$ is used to depict the strength
of the quark pair creation from the vacuum. However, in the
flux-tube model, $\gamma$ is not a constant and is given by
\cite{Kokoski:1985is}
\begin{equation}\label{equa11}
\gamma(\bar{r},\bar{w})=\gamma_0 e^{-\frac{1}{2}bw^2_{\textrm{min}}}.
\end{equation}
}
In Eq.
\eqref{equa9}, $\gamma_0$ is a new phenomenological parameter,
which can be fixed as 14.8 by fitting the experimental data from the PDG
(see Table \ref{table2}). $b$ is the string tension which has the
typical value 0.18 GeV$^2$, and $w_{\textrm{min}}$ is the shortest
distance between the points, where the quark-antiquark pair is created
from the vacuum to the segment connecting the original quark and antiquark in
an initial state (see Fig. \ref{fig2}). The expression of $w_{\textrm{min}}^2$ reads
\begin{equation}\label{equa12}
w_{\textrm{min}}^2 = \left\{ \begin{array}{ll}
w^2 \textrm{sin}^2\theta                & \textrm{if $r\geq w\mid \textrm{cos}\theta\mid $}\\
r^2+w^2-2rw\mid \textrm{cos}\theta\mid & \textrm{if $r< w\mid \textrm{cos}\theta\mid $}\\
\end{array} \right. .
\end{equation}
Then, one can get
the decay width
\begin{eqnarray}\label{equa13}
\Gamma=\frac{\pi}{4}\frac{|\textbf{P}|}{M_A^2}\sum_{LS}|\mathcal{M}^{LS}|^2.
\end{eqnarray}
In order to simplify the calculation,
we use the simple harmonic oscillator (SHO) wave function to depict the meson,
which reads
\begin{equation}\label{equa14}
\psi_{nLM_{L}}(\mathbf{r})=R_{nL}^{\textrm{SHO}}(r)Y_{LM_{L}}(\mathbf{\Omega}_{r})
\end{equation}
with the radial wave function
\begin{equation}\label{equa15}
\begin{array}{c}
R_{nL}^{\textrm{SHO}}(r)=\frac{1}{R^{3/2}}\sqrt{\frac{2n!}{\Gamma(n+L+3/2)}}\biggl(r/R\biggr)^L\\
\times e^{-\frac{r^2}{2R^2}} L_n^{L+1/2}(r^2/R^2).
\end{array}
\end{equation}
Here, $L_n^{L+1/2}(r^2/R^2)$ is an associated Laguerre polynomial.
The parameter $R$ is determined by reproducing the realistic root
mean square radius by solving the Schr\"{o}dinger equation with
the linear potential plus color Coulomb and Gaussian-smeared contact
hyperfine term \cite{Close:2005se}. The $R$ value can be obtained
through the relation \cite{Song:2015nia,Godfrey:2013aaa}
\begin{equation}\label{equa16}
\int |\psi_{nLM_{L}}^{\rm SHO}(\mathbf{r})|^2r^2\,\,d^3\mathbf{r} = \int
|\Phi(\mathbf{r})|^2 r^2\,\,d^3\mathbf{r}.
\end{equation}
The $\Phi(\mathbf{r})$ is the wave function of a certain meson in
the potential model \cite{Godfrey:1985xj}.

\begin{table}[hpb]
\caption{The partial decay widths measured for ten decay channels and the comparison with theoretical calculations.}\label{table2}
\begin{tabular}{ccc}
\toprule [0.4 pt]
\hline
\hline
Decay channel & Experiment (MeV) \cite{Olive:2016xmw}       & Our fit (MeV)\\
\hline
$\rho\rightarrow \pi \pi$              & 147.8 & 83.3  \\
$b_1(1235)\rightarrow \omega \pi$      & 142   & 129.5 \\
$f_2(1270)\rightarrow \pi \pi$         & 156.5 & 85.1  \\
$f_2^\prime\rightarrow K \bar{K}$ & 64.8  & 90.6  \\
$K^\ast\rightarrow K \pi$              & 50.8  & 34.9  \\
$K_0^\ast(1430)\rightarrow K \pi$      & 251   & 381.3 \\
$K_2^\ast(1430)\rightarrow K \pi$      & 49.1  & 63.9  \\
$K_2^\ast(1430)\rightarrow K^\ast \pi$ & 24.3  & 33.8  \\
$K_3^\ast(1780)\rightarrow K^\ast \pi$ & 49.3  & 42.6  \\
$K_3^\ast(1780)\rightarrow K \pi$      & 28.6  & 46.5  \\
\hline
\hline
%\toprule [0.4 pt]
\end{tabular}
\end{table}

\subsection{Fourth pseudoscalar meson nonet}\label{subB}

As discussed in Sec. \ref{sec3}, $\eta(2010)$, $\eta(2100)$, $X(2120)$, and
$\eta(2190)$ can be regarded as the candidates of the third radial
excitation of $\eta(548)$, while $\eta(2225)$ is the candidate of
$\eta^\prime(4S)$. Besides, $\pi(2070)$ can be a $4^1S_0$ state. We
also analyze the mass of the third radial excitation of the kaon, which is
around 2151 MeV and labeled as $K(2150)$ here. In Tables \ref{table3} and
\ref{table4}, the decay channels are listed. In the following, we present the strong decay properties of
these particles.

\begin{figure*}[htbp]
\begin{center}
\includegraphics[width=1\textwidth,scale=1]{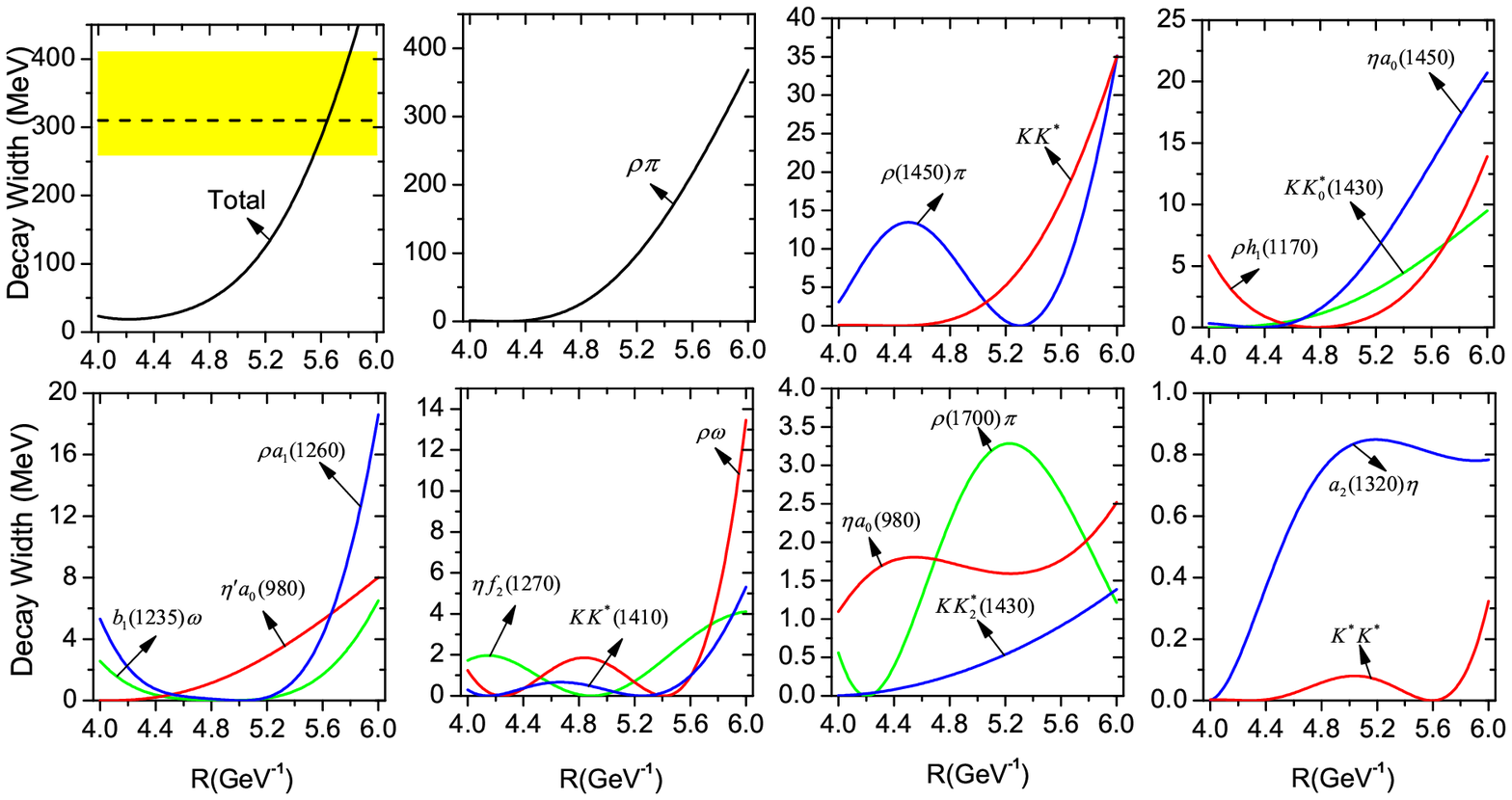}
\caption{The $R$ dependence of two-body strong decay widths of $\pi(2070)$ as a $\pi(4S)$ state. The experimental data are marked by the yellow band. Some tiny channels are not drawn. }\label{fig3}
\end{center}
\end{figure*}

In Fig. \ref{fig3}, we show the total and partial decay widths of
$\pi(2070)$ as a $4^1S_0$ state. By comparing our theoretical results with the experimental data,
the $R$ value lies in the range $5.55\sim5.81 {\rm GeV^{-1}}$, which is
consistent with that in Ref. \cite{Anisovich:2005dt}. $\rho \pi$ is
the dominant decay channel with the width 233 MeV. Here, we choose the
typical value of $R$ as 5.65 GeV$^{-1}$, by which the
center value of the experimental data can be reproduced. $KK^*$ and $\eta a_0(1450)$
are two other sizable decay modes, with the widths 18.22 and 14.38 MeV,
respectively. The partial widths of $\rho(1450)\pi$, $\rho
a_1(1260)$, and $\rho \omega$ are very sensitive to the $R$ value
due to the node effects.

\begin{figure*}[hbp]
\includegraphics[width=0.8\textwidth,scale=0.42]{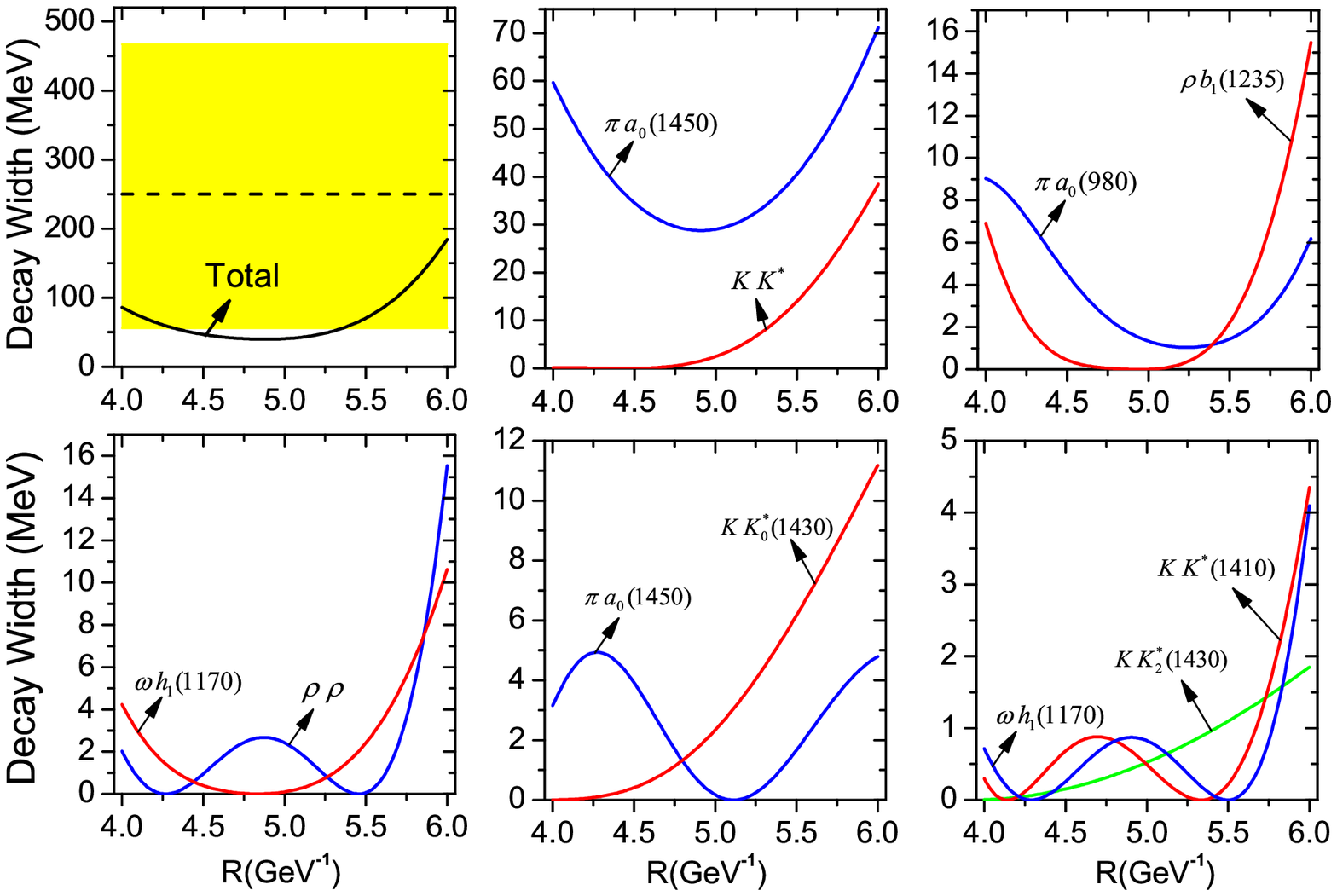}
\caption{The $R$ dependence of the total decay width and partial two-body decay widths of $X(2100)$ as the third radial excitation of $\eta$. The experimental data are marked by the yellow band. Some tiny channels are not drawn. Here, the mixing angle we take is $-2.61^\circ$.}\label{fig4}
\end{figure*}

The $R$ dependence of the decay width of $X(2100)$ is shown in Fig.
\ref{fig4}. We cannot conclude whether or not $X(2100)$ is the
$\eta(4S)$, since the error of the experimental width is too large. From
Fig. \ref{fig4}, we can see that $\pi a_0(1450)$ is the dominant
channel. So we suggest further experiments to study the property
of $X(2100)$ via a $\pi a_0(1450)$ decay mode. In addition, we also
study the strong decay of $\eta(2010)$, $\eta(2100)$, $X(2120)$, and
$\eta(2190)$ under the assignment of the third excitation of
$\eta(548)$. Our results indicate that $\eta(2010)$, $\eta(2100)$,
and $\eta(2190)$ as the $4^1S_0$ isoscalar states are unfavored,
whereas $X(2120)$ seems plausible as as a candidate of the $4^1S_0$ isoscalar state.

\begin{figure*}[hbp]
\centering
\includegraphics[width=0.7\textwidth,scale=0.40]{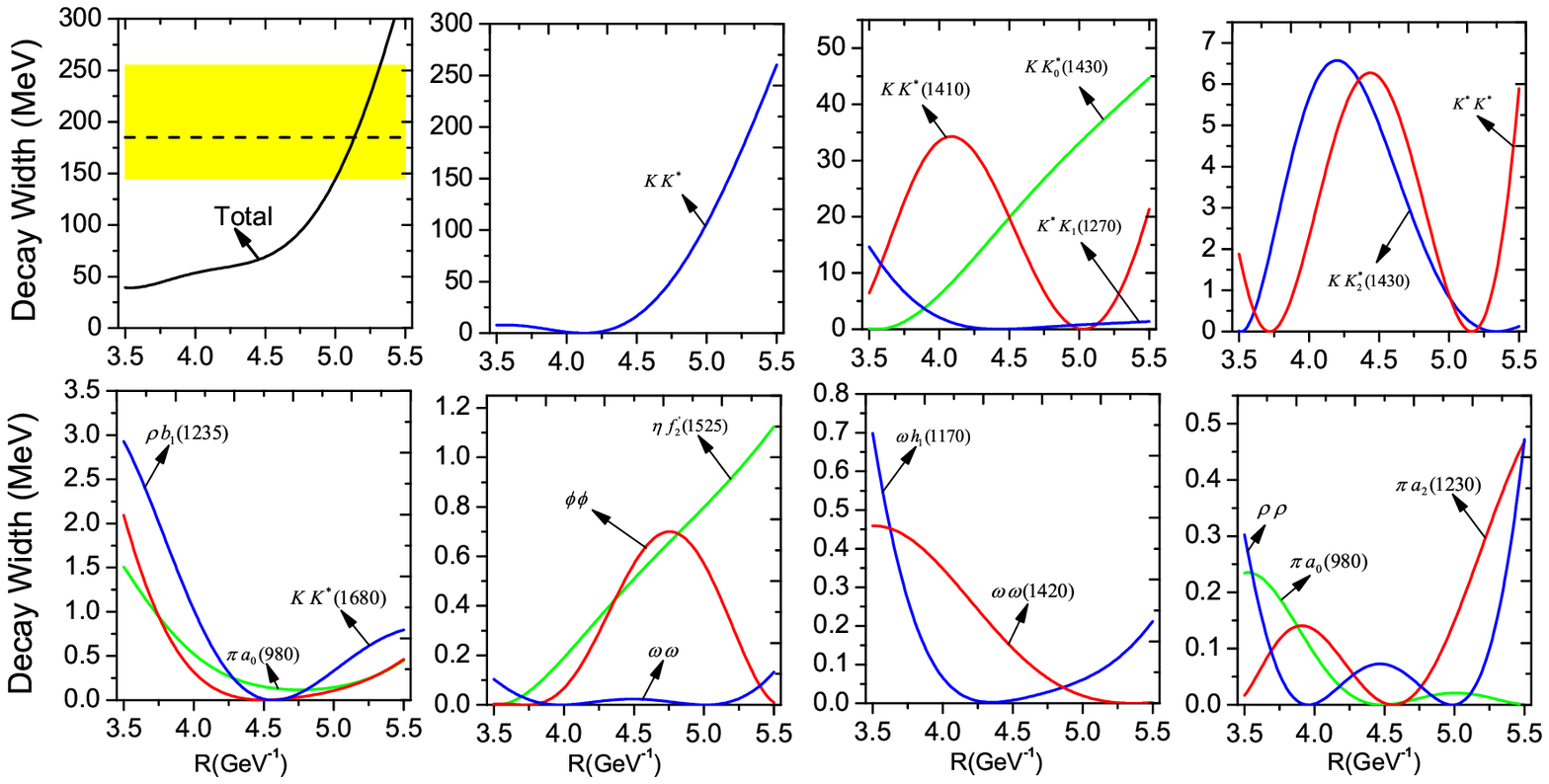}
\caption{The $R$ dependence of total decay width and partial two-body decay widths of $\eta(2225)$ as the third radial excitation of $\eta$. The experimental data are marked by the yellow band. Some tiny channels are not drawn. Here, the mixing angle we take is $6.93^\circ$. }\label{fig5}
\end{figure*}

As mentioned above, $\eta(2225)$ is a good candidate of
$\eta^\prime(4S)$. The plots of decay widths as functions of $R$ are
shown in Fig. \ref{fig5}. The $R$ value is between 5.01
and 5.32 GeV$^{-1}$, which gives an overlap of theoretical and
experimental data. The plausible range of $R$ agrees with that in
Ref. \cite{Chen:2011kp}. The main decay channels are $KK^*$ and
$KK^*_0(1430)$, which have the partial widths of 147 and 36.67 MeV, respectively.
$KK^*(1410)$, $KK^*_2(1430)$ and $K^*K^*$ are highly suppressed due
to node effects. The width of the $\phi\phi$ channel is not ignorable,
which can naturally explain why $\eta(2225)$ is observed by BES via
$J/\psi\to \gamma \phi\phi$ \cite{Ablikim:2008ac}.

\begin{figure*}[hbp]
\begin{center}
\includegraphics[width=1\textwidth,scale=0.42]{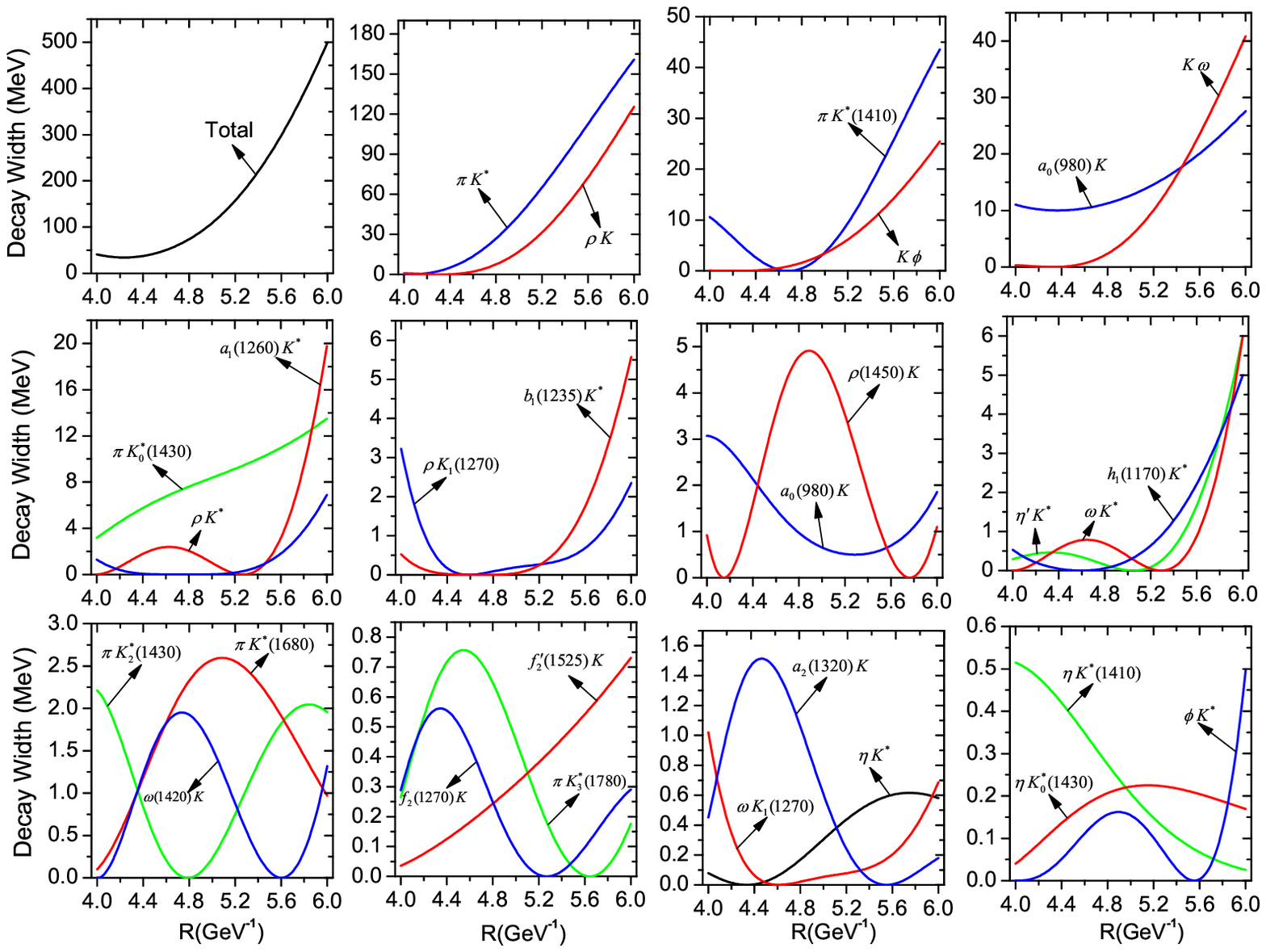}
\caption{The $R$ dependence of two-body decay widths of $K(2150)$ as the third radial excitation of $K$. Some tiny channels are not drawn. }\label{fig6}
\end{center}
\end{figure*}

For $K(2150)$ with $4^1S_0$, there is no experimental information at
present. Its mass is about 2151 MeV. In Fig. \ref{fig6}, we show the strong decay width
of this state, from which we see that the dominant modes are $\pi
K^*$ and $\rho K$. We suggest the experiments to search for
$K(2150)$ via these decay channels.

\begin{table*}[htp]
\caption{The allowed two-body decays of $X(2100),\eta(2225),X(2370),X(2500),\pi(2070)$, and $\pi(2360)$ are marked by $\surd$. Here, $\rho,\,\phi$, and $\omega$ denote  $\rho(770),\,\phi(1020)$, and $\omega(782)$, respectively. }\label{table3}
\centering
%\newsavebox{/tablebox}
%\begin{lrbox}{\tablebox}
\begin{tabular}{m{2cm}m{2.5cm}m{1.6cm}m{1.6cm}m{1.6cm}m{1.6cm}m{2.5cm}m{1.6cm}m{1.6cm}}
%\begin{tabular}{|p{2cm}|p{2cm}|p{2cm}|p{2cm}|p{2cm}|p{2cm}|p{2cm}|p{2cm}|p{2cm}|}
\toprule [0.4 pt]
\hline
\hline
Modes     & Channel               & $X(2100)$ & $\eta(2225)$ & $X(2370)$ & $X(2500)$ & Channel                   & $\pi(2070)$ & $\pi(2360)$\\
\hline
$0^-+0^+$ & $\pi\;a_0(980)$       & $\surd$   & $\surd$      & $\surd$   & $\surd$   & $\eta \; a_0(980)$        & $\surd$     & $\surd$\\
          & $\pi\;a_0(1450)$      & $\surd$   & $\surd$      & $\surd$   & $\surd$   & $\eta \; a_0(1450)$       & $\surd$     & $\surd$\\
          & $\pi(1300)\;a_0(980)$ &           &              & $\surd$   & $\surd$   & $\eta^{\prime}\;a_0(980)$ & $\surd$     & $\surd$\\
          & $K\;K_0^\ast(1430)$   & $\surd$   & $\surd$      & $\surd$   & $\surd$   & $\eta(1295)\;a_0(980)$    &             & $\surd$\\
          &                       &           &              &           &           & $K\;K_0^\ast(1430)$       & $\surd$     & $\surd$\\
%\hline
$1^-+1^+$ & $\rho\;b_1(1235)$     & $\surd$   & $\surd$      & $\surd$   & $\surd$   & $\rho(770)\;h_1(1170)$    & $\surd$     & $\surd$\\
          & $\omega\;h_1(1170)$   & $\surd$   & $\surd$      & $\surd$   & $\surd$   & $\rho(770)\;h_1(1380)$    &             & $\surd$\\
          & $\phi\;h_1(1170)$     &           & $\surd$      & $\surd$   & $\surd$   & $\omega(782)\;b_1(1235)$  & $\surd$     & $\surd$\\
          & $\phi\;h_1(1380)$     &           &              &           & $\surd$   & $K^\ast\;K_1(1270)$       &             & $\surd$\\
          & $K^\ast\;K_1(1270)$   &           &              & $\surd$   & $\surd$   & $K^\ast\;K_1(1400)$       &             & $\surd$\\
          & $K^\ast\;K_1(1400)$   &           &              & $\surd$   & $\surd$   & $\rho(770)\;a_1(1260)$    & $\surd$     & $\surd$\\
%\hline
$0^-+1^-$ & $K\;K^\ast(1410)$     & $\surd$   & $\surd$      & $\surd$   & $\surd$   & $K\;K^\ast$               & $\surd$     & $\surd$\\
          & $K\;K^\ast$           & $\surd$   & $\surd$      & $\surd$   & $\surd$   & $K\;K^\ast(1410)$         & $\surd$     & $\surd$\\
          & $K\;K^\ast(1680)$     &           & $\surd$      & $\surd$   & $\surd$   & $K\;K^\ast(1680)$         &             & $\surd$\\
          & $K(1460)\;K^\ast$     &           &              & $\surd$   & $\surd$   & $K(1460)\;K^\ast$         &             & $\surd$\\
          &                       &           &              &           &           & $\pi\;\rho(770)$          & $\surd$     & $\surd$\\
          &                       &           &              &           &           & $\pi\;\rho(1450)$         & $\surd$     & $\surd$\\
          &                       &           &              &           &           & $\pi\;\rho(1700)$         & $\surd$     & $\surd$\\
          &                       &           &              &           &           & $\pi(1300)\;\rho(770)$    &             & $\surd$\\
%\hline
$0^++1^+$ & $a_0(980)\;a_1(1260)$ &           &              & $\surd$   & $\surd$   & $a_0(980)\;f_1(1285)$     &             & $\surd$\\
          &                       &           &              &           &           & $a_0(980)\;b_1(1235)$     &             & $\surd$\\
%\hline
$1^-+1^-$ & $\rho\;\rho$          & $\surd$   & $\surd$      & $\surd$   & $\surd$   & $\rho(770)\;\omega(782)$  & $\surd$     & $\surd$\\
          & $\rho\;\rho(1450)$    &           & $\surd$      & $\surd$   & $\surd$   & $\rho(770)\;\omega(1420)$ & $\surd$     & $\surd$\\
          & $\omega\;\omega$      & $\surd$   & $\surd$      & $\surd$   & $\surd$   & $\omega(782)\;\rho(1450)$ &             & $\surd$\\
          & $\omega\;\omega(1420)$&           & $\surd$      & $\surd$   & $\surd$   & $K^\ast\;K^\ast$          & $\surd$     & $\surd$\\
          & $\phi\;\phi$          &  $\surd$  & $\surd$      & $\surd$   & $\surd$   & $K^\ast\;K^\ast(1410)$    &             & $\surd$\\
          & $K^\ast\;K^\ast$      & $\surd$   & $\surd$      & $\surd$   & $\surd$   &                           &             &        \\
          & $K^\ast\;K^\ast(1410)$&           &              & $\surd$   & $\surd$   &                           &             &        \\
%\hline
$0^-+2^+$ & $\pi\;a_2(1320)$      & $\surd$   & $\surd$      & $\surd$   & $\surd$   & $\pi\;f_2(1270)$          & $\surd$     & $\surd$\\
          & $\pi\;a_2(1700)$      & $\surd$   & $\surd$      & $\surd$   & $\surd$   & $\eta\;a_2(1320)$         & $\surd$     & $\surd$\\
          & $\eta\;f_2(1270)$     & $\surd$   & $\surd$      & $\surd$   & $\surd$   & $\eta\;a_2(1700)$         &             & $\surd$\\
        & $\eta\;f_2^\prime(1270)$&           & $\surd$      & $\surd$   & $\surd$   & $\eta^\prime\;a_2(1320)$  &             & $\surd$\\
        & $\eta^\prime\;f_2(1270)$&           &              & $\surd$   & $\surd$   & $K\;K_2^\ast(1430)$       & $\surd$     & $\surd$\\
          & $K\;K_2^\ast(1430)$   & $\surd$   & $\surd$      & $\surd$   & $\surd$   &                           &             &        \\
%\hline
$1^-+2^+$&$K^\ast\;K_2^\ast(1430)$&           &              & $\surd$   & $\surd$   & $K^\ast\;K_2^\ast(1430)$  &             & $\surd$\\
          &                       &           &              &           &           & $\rho(770)\;a_2(1320)$    &             & $\surd$\\
%\hline
$0^-+3^-$ & $K\;K_3^\ast(1780)$   &           &              & $\surd$   & $\surd$   & $K\;K_3^\ast(1780)$       &             & $\surd$\\
%\hline
$0^-+4^+$ & $\pi\;a_4(2030)$      &           & $\surd$      & $\surd$   & $\surd$   &                           &             &        \\
%\toprule [0.4 pt]

\hline
\hline
\end{tabular}
%\end{lrbox}%\scalebox{0.87}{\usebox{\tablebox}}
\end{table*}

\begin{table*}[htpb]
\caption{The allowed two-body decays of $K(2150)$ and $K(2414)$ are marked by $\surd$. Here, $\rho,\,\phi$, and $\omega$ denote  $\rho(770),\,\phi(1020)$, and $\omega(782)$, respectively.}\label{table4}
\centering
\begin{tabular}{m{3cm}m{2.8cm}m{2.8cm}m{3cm}m{2.8cm}m{2.8cm}}
\toprule [0.4 pt]
\hline
\hline
Channel                     & $K(2150)$ & $K(2414)$ & Channel                    & $K(2150)$ & $K(2414)$\\
\hline
$\pi\; K_0^\ast(1430)$      & $\surd$   & $\surd$   & $\eta \;K_0^\ast(1430)$    & $\surd$   & $\surd$\\
$\eta^\ast\;K_0^\ast(1430)$ &           & $\surd$   & $K\;a_0(980)$              & $\surd$   & $\surd$\\
$K\;a_0(1450)$              &           & $\surd$   & $\rho\;K_1(1270)$          & $\surd$   & $\surd$\\
$\rho\;K_1(1400)$           &           & $\surd$   & $\omega\;K_1(1270)$        & $\surd$   & $\surd$\\
$\omega\;K_1(1400)$         &           & $\surd$   & $\phi\;K_1(1270)$          &           & $\surd$\\
$K^\ast\;h_1(1170)$         & $\surd$   & $\surd$   & $K^\ast\;h_1(1380)$        &           & $\surd$\\
$K^\ast\;b_1(1235)$         & $\surd$   & $\surd$   & $K^\ast\;a_1(1260)$        &           & $\surd$\\
$K^\ast\;f_1(1285)$         &           & $\surd$   & $K^\ast\;f_1(1420)$        &           & $\surd$\\
$\pi\;K^\ast$               & $\surd$   & $\surd$   & $\pi\;K^\ast(1410)$        & $\surd$   & $\surd$\\
$\pi\;K^\ast(1680)$         &           & $\surd$   & $\eta\;K^\ast$             & $\surd$   & $\surd$\\
$\eta\;K^\ast(1410)$        & $\surd$   & $\surd$   & $\eta\;K^\ast(1680)$       &           & $\surd$\\
$\eta^\prime\;K^\ast$       & $\surd$   & $\surd$   & $\eta^\prime\;K^\ast(1410)$& $\surd$   & $\surd$\\
$\pi(1300)\;K^\ast$         &           & $\surd$   & $K\;\rho$                  & $\surd$   & $\surd$\\
$K\;\omega$                 & $\surd$   & $\surd$   & $K\;\phi$                  & $\surd$   & $\surd$\\
$K\;\omega(1420)$           & $\surd$   & $\surd$   & $K\;\rho(1450)$            & $\surd$   & $\surd$\\
$K\;\omega(1650)$           & $\surd$   & $\surd$   & $K\;\phi(1680)$            &           & $\surd$\\
$K(1460)\;\rho$             &           & $\surd$   & $K(1460)\;\omega$          &           & $\surd$\\
$a_0(980)\;K_1(1270)$       &           & $\surd$   & $a_0(980)\;K_1(1400)$      &           & $\surd$\\
$\rho\;K^\ast$              & $\surd$   & $\surd$   & $\rho\;K^\ast(1410)$       &           & $\surd$\\
$\omega\;K^\ast$            & $\surd$   & $\surd$   & $\omega\;K^\ast(1410)$     &           & $\surd$\\
$\phi\;K^\ast$              & $\surd$   & $\surd$   & $\rho(1450)\;K^\ast$       &           & $\surd$\\
$\omega(1420)\;K^\ast$      &           & $\surd$   & $K\;f_2(1270)$             & $\surd$   & $\surd$\\
$K\;a_2(1320)$              & $\surd$   & $\surd$   & $K\;f_2^\prime(1525)$      & $\surd$   & $\surd$\\
$K\;a_2(1700)$              &           & $\surd$   & $\pi\;K_2^\ast(1430)$      & $\surd$   & $\surd$\\
$\eta\;K_2^\ast(1430)$      & $\surd$   & $\surd$   & $\eta^\prime\;K_2^\ast(1430)$&         & $\surd$\\
$K^\ast\;f_2(1270)$         &           & $\surd$   & $K^\ast\;a_2(1320)$        &           & $\surd$\\
$\omega\;K_2^\ast(1430)$    &           & $\surd$   & $\rho\;K_2^\ast(1430)$     &           & $\surd$\\
$\pi\;K_3^\ast(1680)$       &           & $\surd$   & $\eta\;K_3^\ast(1680)$     &           & $\surd$\\
\hline
\hline
%\toprule [0.4 pt]
\end{tabular}
\end{table*}

\subsection{Fifth pseudoscalar meson nonet}\label{subC}

In the following, we will study the strong decay of the fourth
pseudoscalar meson { nonet}. In Tables \ref{table3} and \ref{table4}, the OZI-allowed decay
channels are listed.

\begin{figure*}[htbp]
\begin{center}
\includegraphics[width=1\textwidth,scale=0.42]{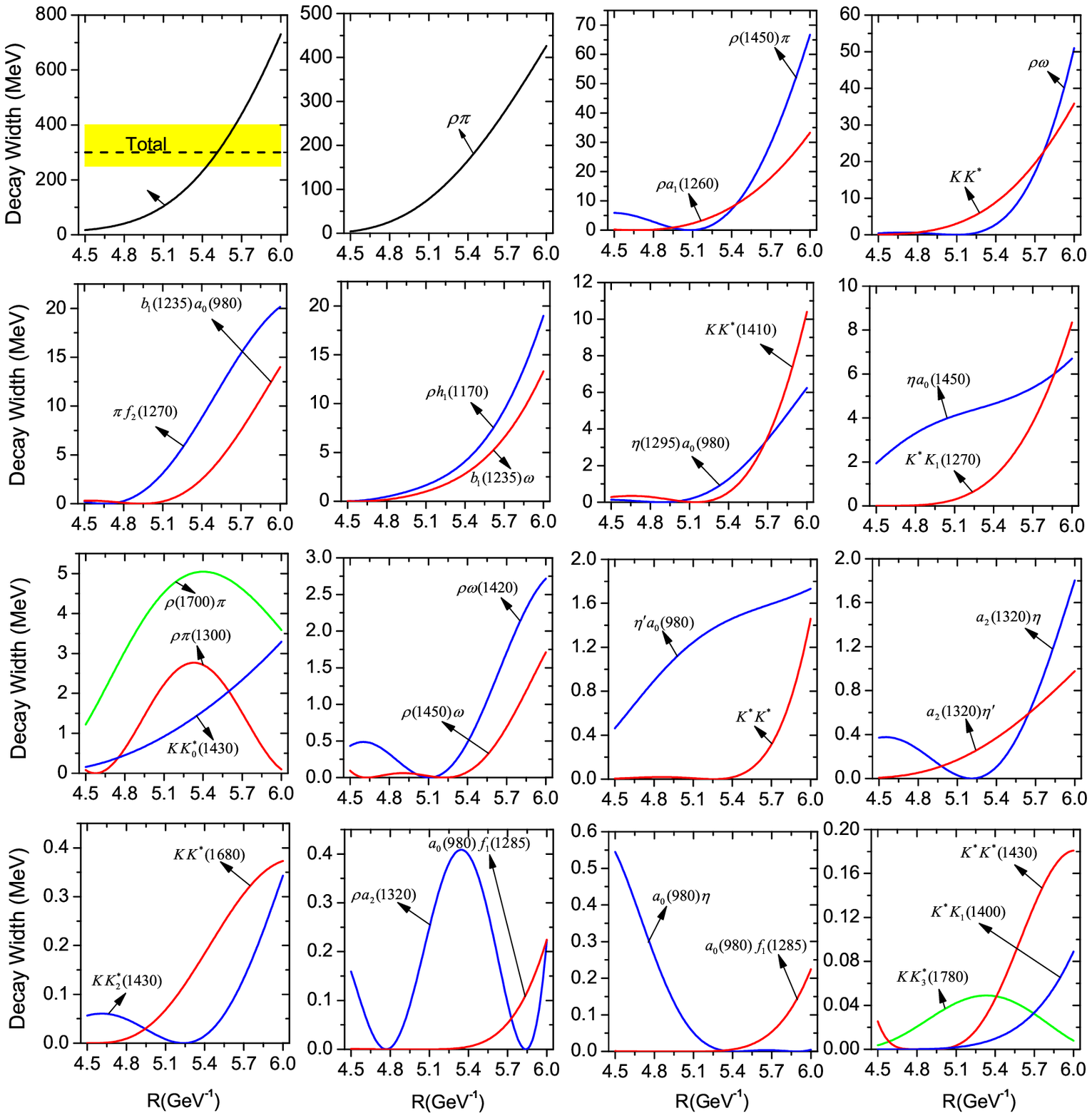}
\caption{The $R$ dependence of two-body strong decay widths of $\pi(2360)$ as a $\pi(5S)$ state. The experimental data are marked by the yellow band. Some tiny channels are not drawn. }\label{fig7}
\end{center}
\end{figure*}

$\pi(2360)$ is a good candidate of the $5^1S_0$ state. In Fig.
\ref{fig7}, we plot the decay width of $\pi(2360)$ depending on $R$.
Comparing to the experimental data, we get the value of $R$ lying in
the range $5.43\sim 5.65\;{\rm GeV^{-1}}$, which is in good agreement
with that in Ref. \cite{Anisovich:2005dt}. Its dominant decay
channel is $\rho\pi$ with the width 205 MeV. In addition, other channels such
as $\rho(1450)\pi$, $KK^*$, $\rho a_1(1260)$, and $\pi f_2(1270)$ are
also important.

\begin{figure*}[htbp]
\centering
\includegraphics[width=1\textwidth,scale=0.42]{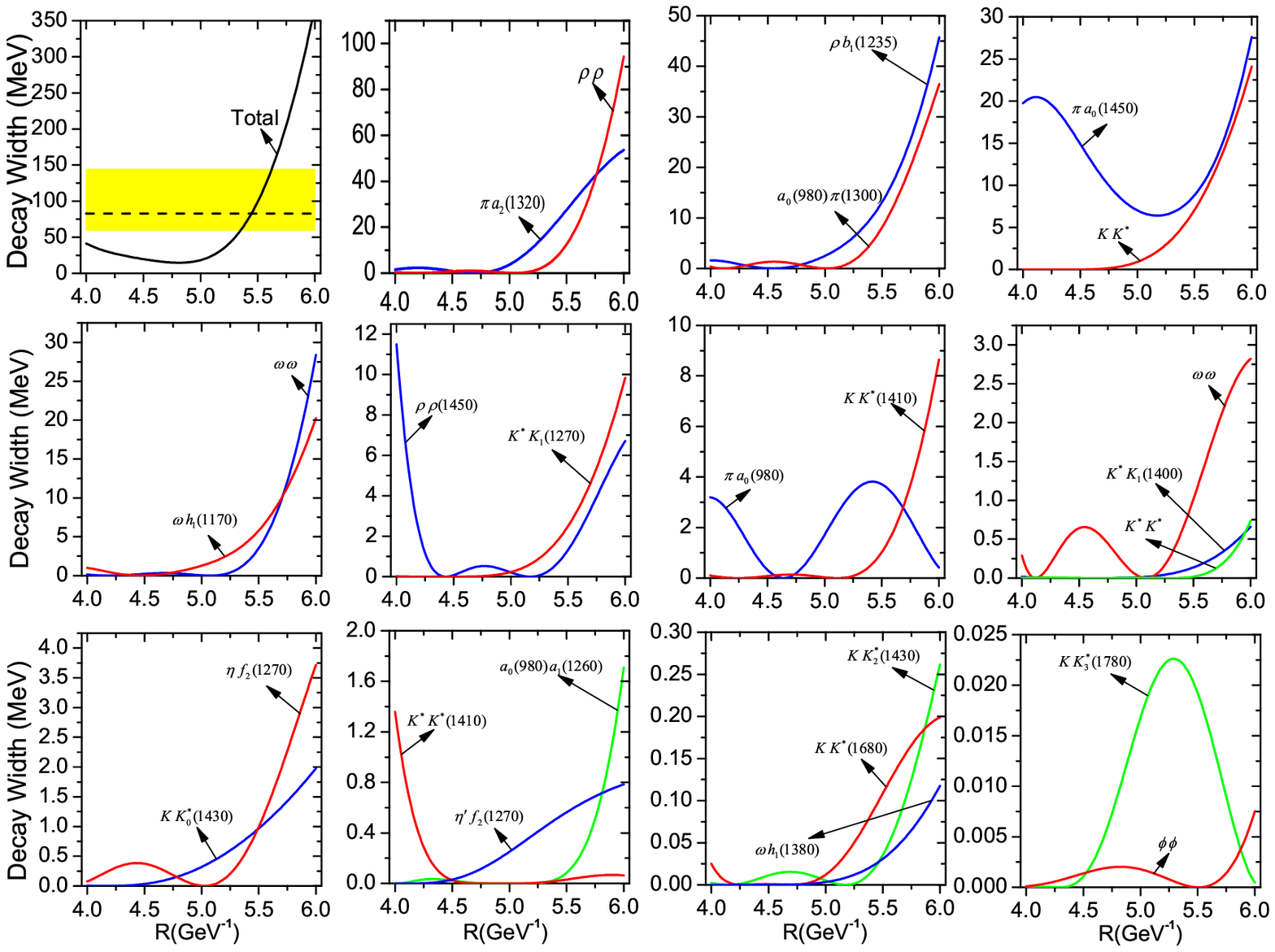}
\caption{The $R$ dependence of total decay widths and the partial two-body decay width of $X(2370)$ as the fourth radial excitation of $\eta$. The experimental data are marked by the yellow band. Some tiny channels are not drawn. Here, the mixing angle we take is $4.18^\circ$. }\label{fig8}
\end{figure*}

The Regge trajectory analysis shows that $X(2370)$ and $\eta(2320)$
can be candidates of the fourth radial excitation of $\eta(548)$.
However, our calculation demonstrates that $\eta(2320)$ cannot be
$\eta(5S)$ since we cannot reproduce the experimental width of $\eta(2320)$ under this assignment. Under the
assignment of $\eta(5S)$, we can
 get the width of $X(2370)$ which is shown
in Fig. \ref{fig8}. If choosing $R$ around 5.44 GeV$^{-1}$ which is
similar to Ref. \cite{Yu:2011ta}, the theoretical value of the total
width is equal to the experimental central value. From Fig.
\ref{fig8}, we can see that the $\rho\rho$, $KK^*$, $\pi a_0(1450)$,
$a_0(980)\pi(1300)$, $\pi a_2(1320)$, and $\rho b_1(1235)$ channels
are important.

\begin{figure*}[htbp]
\centering
\includegraphics[width=1\textwidth,scale=0.42]{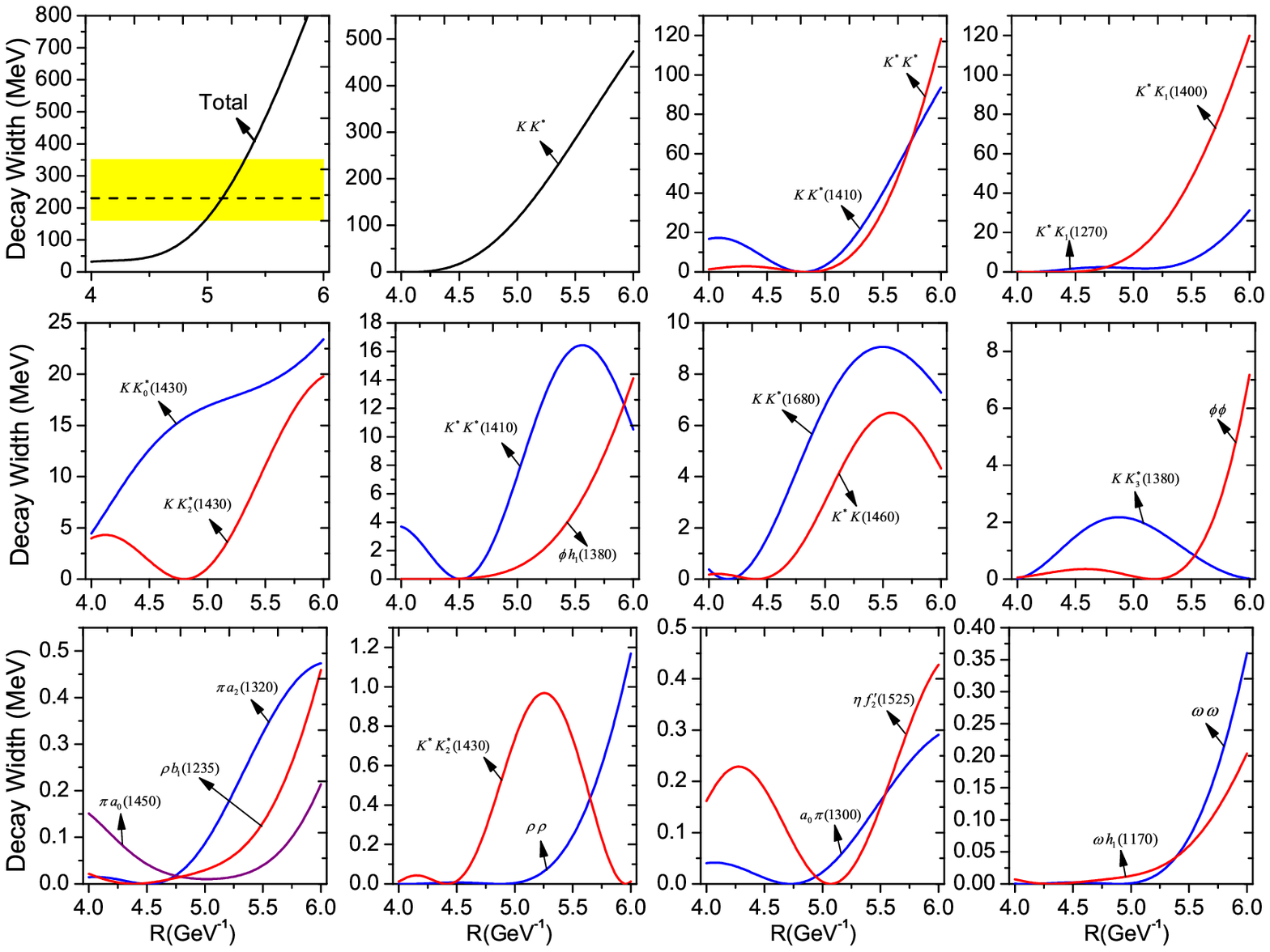}
\caption{The $R$ dependence of total decay widths and the partial two-body decay width of $X(2500)$ as the fourth radial excitation of $\eta^\prime$. The experimental data are marked by the yellow band. Some tiny channels are not drawn. Here, the mixing angle we take is $4.18^\circ$. }\label{fig9}
\end{figure*}

According to the mass spectrum analysis, $X(2500)$ is a good
candidate of $\eta^\prime(5S)$. In Fig. \ref{fig9}, we plot the
decay width of $X(2500)$ under the assignment of the fourth radial
excitation of $\eta^\prime(958)$. The value of $R$ corresponding to the
central value of the experimental width falls in the range of $4.98\sim
5.32\;{\rm GeV^{-1}}$. Choosing a typical value of $R$ as 5.13
GeV$^{-1}$, the dominant decay { mode $KK^{\ast}$ has} the width of
154 MeV. Besides, the $\phi\phi$ channel is not ignorable, which can
explain why $X(2500)$ is observed in the $\phi\phi$ channel.

\begin{figure*}[hbbp]
\begin{center}
\includegraphics[width=0.85 \textwidth,scale=0.15]{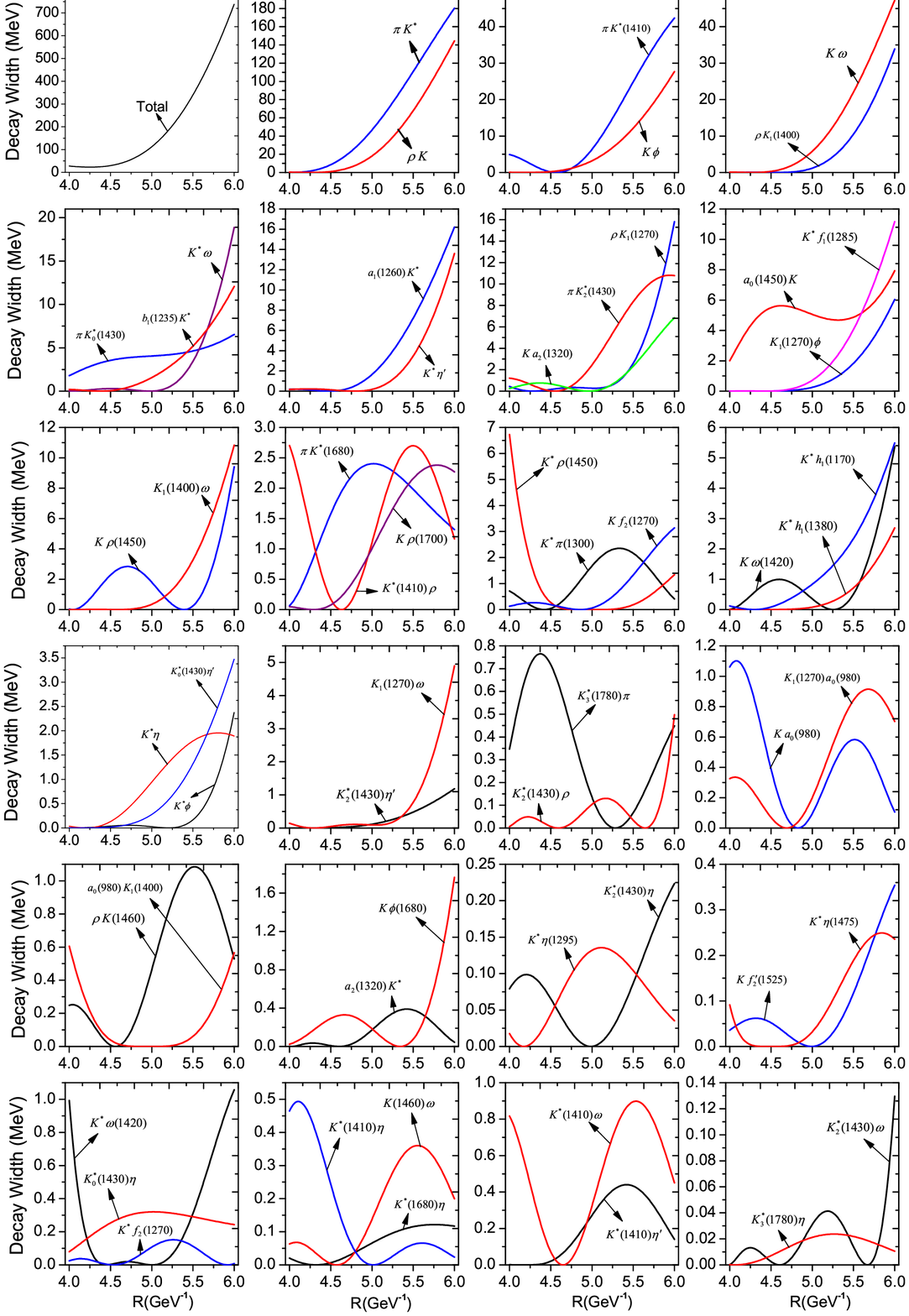}
\caption{The $R$ dependence of two-body decay widths of $K(2414)$ as the fourth radial excitation of $K$. Some tiny channels are not drawn here. }\label{fig10}
\end{center}
\end{figure*}

As mentioned in Sec. \ref{sec3}, the $5^1S_0$ state of the kaon
labeled by $K(2414)$ has a mass of 2414 MeV. The strong decay, which
is shown in Fig. \ref{fig10}, is dominated by $\pi K^*$ and $\rho K$.
Additionally, $\pi K^*(1410)$, $K\phi$, $K\omega$, and $a_0(980)K$ are also
important. This results will be helpful to explore $K(2414)$ in experiment.

%%%%%%%%%%%%%%%%%%%%%%%%%%%%%%%%%%%%%%%%%%%%%%%%%%%%%%%%%%%%%
\section{conclusions and discussion}\label{sec5}
%%%%%%%%%%%%%%%%%%%%%%%%%%%%%%%%%%%%%%%%%%%%%%%%%%%%%%%%%%%%%
Inspired by the observed $X(2100)$,
%$\eta(2100)$,
$X(2500)$, and $\eta(2225)$,
we have tried to construct new pseudoscalar meson nonets including these states.
$\pi$, $K$, $\eta(548)$, and $\eta^\prime(958)$ belong to the ground
state pseudoscalar nonet. As stated in Ref. \cite{Klempt:2007cp},
$\pi(1300)$, $K(1460)$, $\eta(1295)$, and $\eta(1475)$ form the first
radial excitation of the $0^-$ meson nonet. The $\pi(1800)$, $K(1830)$,
$\eta(1760)$, and $X(1835)$ are grouped into the third pseudoscalar nonet. In
this paper, we have  speculated that the fourth and fifth pseudoscalar
meson nonets are made by $\{\pi(2070), K(2150), \eta(4S),
\eta(2225)\}$ and $\{\pi(2360), K(2414), \eta(5S), \eta(2500)\}$,
respectively. Here, the candidates for $\eta(4S)$ could be
$\eta(2010)$, $\eta(2100)$, $\eta(2190)$, $X(2120)$, and $X(2100)$,
while $\eta(5S)$ could be either $X(2370)$ or $\eta(2320)$. Note
that $K(2414)$ and $K(2150)$ are predicted particles by using
diagonalization of the mass squared matrix and the Gell-Mann-Okubo mass
formula. Our speculation has satisfied the Regge trajectories.

Within this scheme, the strong decay of these states has been studied by
the flux-tube model. $X(2100)$ or $\eta(2100)$ as a $4^1S_0$ state
is undetermined since the experimental information is not
sufficient. The suggested channel of these two states for further
experimental studies is $\pi a_0(1450)$. We exclude $\eta(2010)$ and
$\eta(2190)$ to be the third radial excitation of $\eta(548)$.
$X(2120)$ is a good candidate of $\eta(4S)$, which agrees with the
conclusion in Ref. \cite{Yu:2011ta}. In addition, $\pi(2070)$ and
$\eta(2225)$ can be explained as $\pi(4S)$ and $\eta^\prime(4S)$.
The predicted particle $K(2150)$ is a candidate for $K(4S)$,
the dominant channels $\pi K^*$ and $\rho K$ of which can be tested in
future experiments.

Comparing the theoretical and experimental widths, we find that
the candidate for $\eta(5S)$ cannot be $\eta(2320)$ but $X(2370)$.
The newly observed $X(2500)$ can be interpreted as
$\eta^\prime(5S)$. Moreover, we have studied the strong decay of
$\pi(2360)$ assuming the quantum number is $5^1S_0$, where the
calculated width agrees with the experimental one with $R$ around
$5.51$ GeV$^{-1}$. The predicted strange meson $K(2414)$ with
quantum number $5^1S_0$ has been also studied. The total width is in the
range of 112.1 $\sim$ 371.8 MeV with $R$ in the range of 5.0 $\sim$
5.55 GeV$^{-1}$. We have suggested a further experimental search for this
state via $\pi K^*$ and $\rho K$ channels. {We have summarized
the arrangement of the mesons in Table \ref{table1}}.

\begin{table*}[htpb]
\caption{{ The pseudoscalar nonets predicted in this paper.}}\label{table1}
\begin{tabular}{ccccc}
\toprule [0.4 pt] \hline \hline
$1S$ &  $2S$ & $3S$ & $4S$ & $5S$\\
\hline
$\eta,$& $\eta(1295)$&$\eta(1760)$&$X(2120)/\eta(2100)/X(2100)$&$X(2370)$\\
$\eta^\prime$ & $\eta(1475)$& $X(1835)$& $\eta(2225)$&$X(2500)$\\
$K$ & $K(1460)$ & $K(1830)$& $K(2150)$ &$K(2414)$\\
$\pi$ & $\pi(1300)$ & $\pi(1800)$ &$\pi(2070)$ &$\pi(2360)$\\
\hline \hline %\toprule [0.4 pt]
\end{tabular}
\end{table*}

The important information of pseudoscalar states provided by BESIII greatly enriches our
knowledge on the light hadron spectra. Further experimental and theoretical efforts will be helpful in establishing new pseudoscalar meson nonets.
The predicted behaviors of the discussed states can be tested in the near future, and we would like to have more experimental progress of BESIII and the forthcoming BelleII.

\vfil

%%%%%%%%%%%%%%%%%%%%%%%%%%%%%%%%%%%%%%%%%%%%%%%%%%%%%%%%%%%%%
\section*{Acknowledgments}
We would like to thank the anonymous referee for his suggestions and comments. This  work is supported in part by National Natural Science Foundation of China under the
Grants No. 11222547 and No. 11175073 and the Fundamental Research Funds for the Central Universities. Xiang Liu is also supported by the National Program for Support of
Top-Notch Young Professionals.
%%%%%%%%%%%%%%%%%%%%%%%%%%%%%%%%%%%%%%%%%%%%%%%%%%%%%%%%%%%%%

\end{document}